\documentclass[aps,pra,reprint,groupeaddress]{revtex4-1}

\usepackage[utf8]{inputenc}
\usepackage[english]{babel}
\usepackage[colorlinks=true,linkcolor=blue,citecolor=blue,urlcolor=blue]{hyperref}
\usepackage{amssymb}
\usepackage{amsmath}
\usepackage{amsfonts}
\usepackage{braket}
\usepackage{esdiff}
\usepackage{nicefrac}
\usepackage{graphicx}
\usepackage{xcolor}

\begin{document}


\title{Time-resolved statistics of photon pairs in two-cavity Josephson photonics}



\author{Simon Dambach}
\affiliation{Institute for Complex Quantum Systems and IQST, Ulm University, Albert-Einstein-Allee 11, 89081 Ulm, Germany}
\author{Björn Kubala}
\affiliation{Institute for Complex Quantum Systems and IQST, Ulm University, Albert-Einstein-Allee 11, 89081 Ulm, Germany}
\author{Joachim Ankerhold}
\affiliation{Institute for Complex Quantum Systems and IQST, Ulm University, Albert-Einstein-Allee 11, 89081 Ulm, Germany}


\date{April 22, 2017}

\begin{abstract}
We analyze the creation and emission of pairs of highly nonclassical microwave photons in a setup where a voltage-biased Josephson junction is connected in series to two electromagnetic oscillators. 
Tuning the external voltage such that the Josephson frequency equals the sum of the two mode frequencies, each tunneling Cooper pair creates one additional photon in both of the two oscillators. The time-resolved statistics of photon emission events from the two oscillators is investigated by means of single- and cross-oscillator variants of the second-order correlation function $g^{(2)}(\tau)$ and the waiting-time distribution $w(\tau)$. They provide insight into the strongly correlated quantum dynamics of the two oscillator subsystems and reveal a rich variety of quantum features of light including strong antibunching and the presence of negative values in the Wigner function. 
\end{abstract}


\maketitle


\section{Introduction}

Started by a pioneering experiment at Saclay \cite{Hofheinz2011}, the use of single-charge tunneling events as a new, unconventional source of microwave photons has garnered increased interest. Employing a voltage-biased Josephson junction connected in series to one or several microwave cavities enables a near perfect current-to-light conversion.
The regime of strong nonequilibrium, where stimulated emission processes dominate, is reached in realizations based on a voltage-biased Cooper-pair transistor and a high-$Q$ cavity~\cite{Chen2014}. A related realization incorporating several double quantum dots, while converting less perfectly, nonetheless has been shown to reach a lasing-like state \cite{Liu2015}.
The interaction of a \emph{single charge} (or the tunneling Cooper pair here) and a \emph{single photon} (or pairs of photons in the following) combines  in a \emph{Josephson photonics} setup with the inherent nonlinearity of the junction to create nonconventional microwave light with non-Gaussian or even quantum features.
While considerable attention has already been devoted to understanding the simpler limits of (incoherent) $P(E)$-theory and nonlinear (semi-)classical dynamics and their crossover into a fully quantum regime \cite{Leppaekangas2013,Armour2013,Gramich2013,Kubala2015,Armour2015,Trif2015,Dykman2012,Leppaekangas2015},  powerful signatures of the quantum nature of the device can be found in the time-resolved statistics of the emitted light. 

Here, we extend previous work investigating correlation function and waiting-time distribution for a single cavity \cite{Dambach2015} to a two-cavity scenario. This offers a wider variety of feedback and blocking mechanisms between the junction and the two cavities. The cross-correlated nature of the light emitted from the two cavities can also serve as an important starting point towards issues of bi- or multipartite entanglement.

After introducing model and observables in Sec.~\ref{sec_Model}, we extensively study three interesting simpler scenarios (Secs.~\ref{subsec_Nondegenerate_parametric_amplifier} to \ref{subsec_Anti-Jaynes-Cummings_system}) arising for special choices of parameters from the generic two-cavity case, which is finally investigated in Sec.~\ref{subsec_Crossover_in_the_weak-driving_limit} building on those earlier results.

\begin{figure*}
\centering
\includegraphics[width=1.0\textwidth]{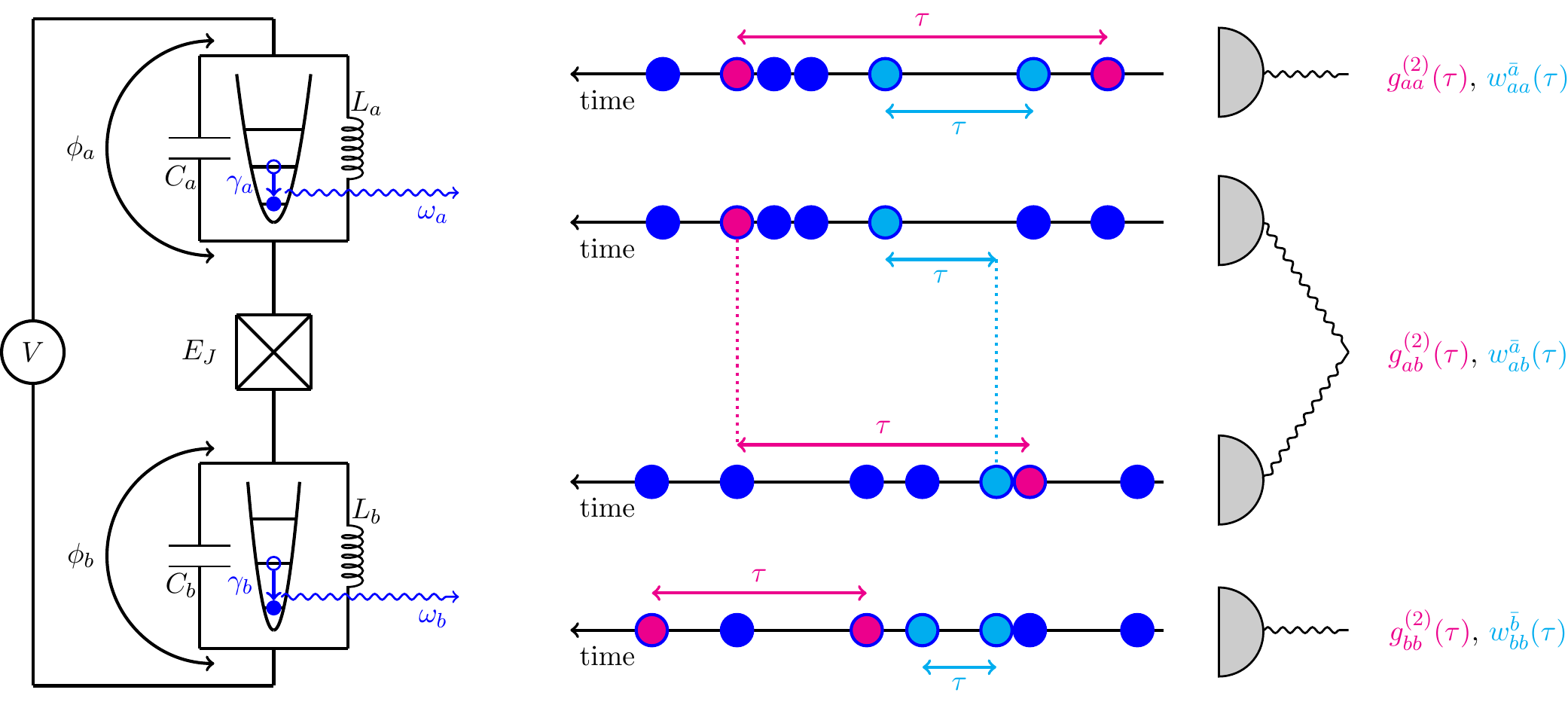}
\caption{Sketch of the effective circuit model and a schematic illustration of the different observables used to study the time-resolved statistics of the emitted microwave radiation. Tuning the voltage to the resonance $\omega_{J}=2eV/\hbar=\omega_{a}+\omega_{b}$, each Cooper pair tunneling across the junction excites two photons, one in each of the two  LC resonators, which subsequently leak out and can be observed. The time-resolved statistics of the photon emission events is investigated studying correlations between two photon emission events. Various correlation functions involving one or both resonators and allowing or excluding intermediate emission events are employed.
}
\label{fig_Fig_1} 
\end{figure*}

\section{Model}
\label{sec_Model}

In this section, we briefly discuss our theoretical modeling consisting of an effective Hamiltonian in rotating-wave approximation (RWA) and a quantum master equation of Lindblad form. On the basis of this master-equation formalism, we introduce different versions of the second-order correlation function $g^{(2)}(\tau)$ and waiting-time distribution $w(\tau)$ allowing for either single-oscillator or cross-oscillator measurements.  

\subsection{Hamiltonian}
\label{subsec_Hamiltonian}

We consider a setup (see Fig.~\ref{fig_Fig_1} for an effective circuit model) consisting of a Josephson junction (JJ) and two series-connected LC oscillators, denoted by a and b, across which an external voltage $V$ is applied. By tuning the voltage, we access that resonance where the Josephson frequency $\omega_{J}=2eV/\hbar$ and the sum of the two mode frequencies $\omega_{a(b)}=1/\sqrt{L_{a(b)}C_{a(b)}}$ matches, i.e., the energy lost by a single Cooper pair tunneling across the junction equals the energy that is necessary for the creation of one photon in each of the two oscillators. To avoid competition with processes in which two photons are created within a single mode, we require the modes here to be nondegenerate, $\omega_{a}\neq\omega_{b}$.   

The system can be described \cite{Armour2015,Trif2015} by the effective Hamiltonian
\begin{equation}
H=\hbar\omega_{a}n_{a}+\hbar\omega_{b}n_{b}-\frac{E_{J}}{2}\left(e^{i\omega_{J}t}e^{i\phi_{a}}e^{i\phi_{b}}+\mathrm{h.c.}\right),
\end{equation}
where the photonic number operator $n_{a}=a^{\dagger}a$ and the phase $\phi_{a}=\sqrt{\kappa_{a}}(a^{\dagger}+a)$ of oscillator a are given in terms of the conventional bosonic creation and annihilation operators $a^{\dagger}$ and $a$ with $[a,a^{\dagger}]=1$ and analogous for oscillator b. The dimensionless parameter $\kappa_{a}=E_{C_{a}}/\hbar\omega_{a}$ characterizes the granularity of charges in the circuit via the charging energy $E_{C_{a}}=2e^{2}/C_{a}$ and is at the same time a measure for the zero-point quantum fluctuations in oscillators a. 

Performing a rotating wave approximation after a unitary transformation to a rotating frame via $U(t)=\exp{[i(\omega_{a}-\Delta_{a})n_{a}t]}\exp{[i(\omega_{b}-\Delta_{b})n_{b}t]}$  yields
\begin{equation}
\begin{aligned}
\!\!\!\!H_{\mathrm{RWA}}\!\!=&\hbar\Delta_{a}n_{a}+\hbar\Delta_{b}n_{b}\\
&\!+\!\frac{\tilde{E}_{J}}{2}\!:\!\left(a^{\dagger}b^{\dagger}\!+\!ab\right)\!\frac{J_{1}\left(\sqrt{4\kappa_{a} n_{a}}\right)\!J_{1}\left(\sqrt{4\kappa_{b} n_{b}}\right)}{\sqrt{n_{a}}\sqrt{n_{b}}}\!:.\!\!
\end{aligned}  
\label{eq_H_RWA}
\end{equation}
Here, $\Delta_{a}=\omega_{a}-\tilde{\omega}_{a}$ denotes the detuning with respect to the resonance condition $\omega_{J}=\tilde{\omega}_{a}+\tilde{\omega}_{b}$ (we assume, however, $\Delta_{a}=\Delta_{b}=0$ in the remainder of the paper) and $\tilde{E}_{J}=E_{J}\exp{[-(\kappa_{a}+\kappa_{b})/2]}$ represents a renormalized Josephson energy. The colons indicate normal ordering.

The inherent nonlinearity of the Josephson junction enters here as a nonlinear driving term in form of a normal-ordered product of Bessel functions of the first kind renormalizing the fundamental creation/annihilation term, $a^{\dagger}b^{\dagger}+ab$. The effect of these nonlinearities on the system's dynamics depend on both the Josephson energy $E_{J}$, which can be interpreted as the driving strength, and the $\kappa_{a(b)}$ parameters, which reflect the charge quantization of the Cooper pair current and crucially influence the transition matrix elements of the drive Hamiltonian between neighboring resonator states (see below, Sec.~\ref{subsec_Transition_matrix_elements}). Experimentally, $E_{J}$ is easily tunable to a certain extent via the magnetic flux when using a SQUID geometry. Notably, the recent development of ultra-low temperature scanning tunneling microscopes \cite{Berthold2016,Ruby2015} has opened up the possibility to vary the Josephson coupling via the tip-sample distance over a much broader range. The $\kappa$ parameter is fixed by design and can only be slightly varied in situ. Earlier experimental realizations were limited to the small-$\kappa$ regime ($\kappa\approx0.1$) of low impedances, however, recent progress already allows to reach values up to $\kappa\approx1.6$. 

\subsection{Quantum master equation}
\label{subsec_Quantum_master_equation}

The finite quality factor $Q_a=\omega_a/\gamma_a$ limiting the lifetime $1/\gamma_a$ of excited photons in resonator a is dominated by photon leakage into microwave output ports. The dynamics of the density operator of the system in the zero-temperature limit can hence be described by a master equation in Lindblad form \cite{Breuer2002}
\begin{equation}
\begin{aligned}
\!\!\!\!\diff{}{t}\rho=\mathfrak{L}\rho
=-\frac{i}{\hbar}\left[H_{\mathrm{RWA}},\rho\right]+&\frac{\gamma_{a}}{2}\!\left(2a\rho a^{\dagger}-n_{a}\rho-\rho n_{a}\right)\\+&\frac{\gamma_{b}}{2}\!\left(2b\rho b^{\dagger}-n_{b}\rho-\rho n_{b}\right),
\end{aligned}
\end{equation}
which can be written in terms of the Liouvillian superoperator $\mathfrak{L}$. The asymmetry of the two decay rates is characterized by $r=\gamma_{a}/\gamma_{b}$. Note that we have not included here the impact of local voltage fluctuations at the JJ into our model, which would enter in form of an additional dissipator associated with rate $\gamma_{J}$ in the master equation above. This is justified since both experimental and theoretical investigations \cite{Hofheinz2011,Gramich2013} show that the voltage noise is weak ($\gamma_{J}\ll\gamma_{a},\gamma_{b}$) and we will neglect its effect on all observables considered here.

The full time evolution of the system's dynamics $\mathfrak{L}=\mathfrak{L}_{\bar{a}\bar{b}}+\mathfrak{J}_{a}+\mathfrak{J}_{b}$ is basically constituted from three parts \cite{Molmer1993,Plenio1998,Brandes2008}. A jump operator $\mathfrak{J}_a$ describes the emission of a photon from oscillator a, $\mathfrak{J}_a\rho=\gamma_{a}a\rho a^{\dagger}$, and the same for b.
The remaining part $\mathfrak{L}_{\bar{a}\bar{b}}$ describes the dissipative but deterministic dynamics during the time interval where no photon emission events occur. On the basis of this decomposition, we define here the time evolution $\mathfrak{L}_{\bar{a}}=\mathfrak{L}-\mathfrak{J}_{a}$ which excludes emission events from oscillator a but allows for those from oscillator b. 
 
\subsection{Observables for studying correlations}
\label{subsec_Observables_for_studying_correlations}

To investigate the time-resolved statistics of the emitted photons, we make use of two different but closely related observables, the \emph{second-order correlation function} $g^{(2)}(\tau)$ and the \emph{waiting-time distribution} (WTD) $w(\tau)$, which are both well-established tools in the field of quantum optics \cite{Mandel1962,Walls1994,Carmichael1989}. 
Resolving the photon emission processes from the two different oscillators allows to introduce different versions of $g^{(2)}(\tau)$ and $w(\tau)$ functions:
\begin{equation}
\,\,\!g_{kl}^{(2)}(\tau)\!=\!\frac{\langle\mathfrak{J}_{k}e^{\mathfrak{L}\tau}\mathfrak{J}_{l}\rangle_{\mathrm{st}}}{\langle\mathfrak{J}_{k}\rangle_{\mathrm{st}}\langle\mathfrak{J}_{l}\rangle_{\mathrm{st}}}
{\!\quad \mbox{and}\!\quad}
w^{\bar{k}}_{kl}(\tau)\!=\!\frac{\langle\mathfrak{J}_{k}e^{\mathfrak{L}_{\bar{k}}\tau}\mathfrak{J}_{l}\rangle_{\mathrm{st}}}{\sqrt{\langle\mathfrak{J}_{k}\rangle_{\mathrm{st}}\langle\mathfrak{J}_{l}\rangle_{\mathrm{st}}}}\!\!
\end{equation}   
with $j,k\in\lbrace a,b\rbrace$. Here, $\langle\dots\rangle_{\mathrm{st}}$ indicates steady state expectation values, $\langle O\rangle_{\mathrm{st}}=\mathrm{Tr}\lbrace O\rho_{\mathrm{st}}\rbrace$, with $\mathfrak{L}\rho_{\mathrm{st}}=0$. 

The $g^{(2)}_{kl}(\tau)$ function measures correlations between an emission event from oscillator k and a prior event from oscillator l separated by a time interval $\tau$. The $w^{\bar{k}}_{kl}(\tau)$ function is the probability distribution to detect a delay $\tau$ between an emission event from oscillator l and the first subsequent detection of a photon from oscillator k, see Fig.~\ref{fig_Fig_1} for illustration. Note the superscript $\bar{k}$ in the WTD highlighting that no jumps from oscillator k are allowed during the interval $\tau$. 

\section{Results}
\label{sec_Results}

The dynamics of the JJ-cavity system is driven by Cooper pairs tunneling across the junction and creating photons in the cavities which in turn act back on the tunneling process \-- stimulating or hindering further emission processes. To understand the resulting complex interplay of the (nonlinear) creation process and backaction, we turn to the heart of the nonlinearity of the Josephson junction. 

\subsection{Transition matrix elements}
\label{subsec_Transition_matrix_elements}

In contrast to many other nonlinear resonator systems, the nonlinearity of the JJ-cavity system stems from a nonlinear driving, not a nonlinear spectrum of excitation energies. It is thus best understood by investigating the transition matrix elements
\begin{equation}
\begin{aligned}
\!\!\!\!T_{m_{a},m_{b},m_{a}\!+\!1,m_{b}\!+\!1}&=
\Braket{m_{b},m_{a}|H_{RWA}|m_{a}\!+\!1,m_{b}\!+\!1}\\
&=T_{m_{a},m_{a}\!+\!1}\; T_{m_{b},m_{b}\!+\!1}
\end{aligned}
\end{equation}
between neighboring occupation states $\ket{m_{a},m_{b}}$ and $\ket{m_{a}+1,m_{b}+1}$ of the cavities a and b.
Due to the obvious factorization into matrix elements involving a single cavity only, the physics of the two-cavity excitation processes carries over many of the features found in the single-mode case \cite{Kubala2015,Dambach2015}. 

There we found that the parameter $\kappa$ crucially determines the impact of nonlinearities. In particular, through a proper choice of $\kappa$, the transition between states $\ket{m}$ and $\ket{m+1}$ could be blocked for some chosen $m$ and, hence, an $m+1$-level system can be engineered. Considering the limit $\kappa \rightarrow 0$, the transition matrix elements of a harmonically-driven oscillator are recovered. Following this line of reasoning, we will in the following consider three special cases, where the two nonlinearly-driven cavities are reduced to simpler systems:
(i) the limit $\kappa_{a},\kappa_{b}\rightarrow 0$ resulting in a nondegenerate parametric amplifier (PA), (ii) the case of two linked two-level systems (TLSs) obtained for $\kappa_{a}=\kappa_{b}=2$,  and (iii) the anti-Jaynes-Cummings (anti-JC) system, coupling a harmonic oscillator and a two-level system, realized by $\kappa_{a}\rightarrow 0$ and $\kappa_{b}=2$.     

\subsection{Nondegenerate parametric amplifier}
\label{subsec_Nondegenerate_parametric_amplifier}

\begin{figure}
\centering\hspace*{-0.75cm}
\includegraphics[width=\columnwidth]{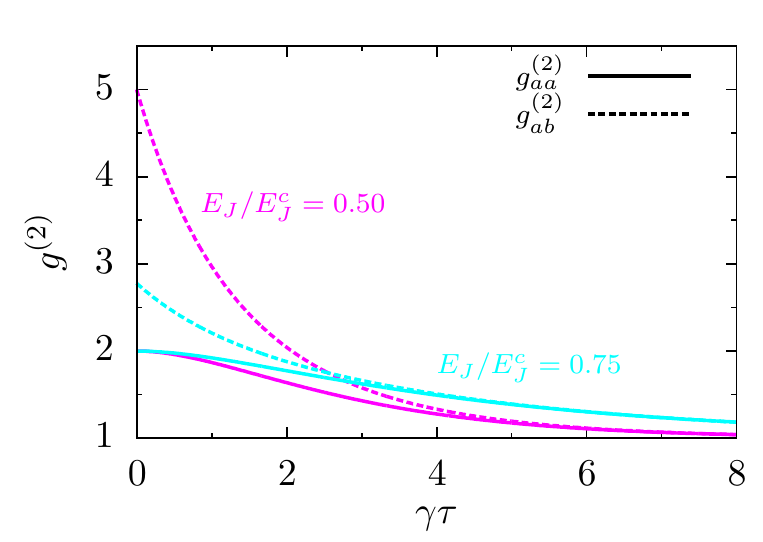}
\caption{Second-order correlation functions $\smash{g^{(2)}_{aa}(\tau)}$ and $\smash{g_{ab}^{(2)}(\tau)}$ in the limit of a nondegenerate parametric amplifier ($\kappa_{a(b)}\rightarrow 0$) in the subthreshold regime, $E_{J}/E^{c}_{J}<1$, for $r=1$. Simultaneous excitation of photons in both resonators results in a bunching for cross- and intra-cavity correlation functions.} 
\label{fig_Fig_2}
\end{figure}

Taking the limit $\kappa_{a},\kappa_{b}\rightarrow 0$ in the Hamiltonian [Eq.~\eqref{eq_H_RWA}], the nonlinear Bessel functions drop from the problem and the well-known case of a nondegenerate parametric amplifier Hamiltonian is recovered. This linearized version of the two-cavity case thus has closer similarities to a single cavity weakly driven at the \emph{two-photon} resonance (which was shown in Ref.~\cite{Dambach2015} to result in a degenerate PA) than to the fundamental \emph{single-photon} resonance of the single cavity (which reduces to the linearly-driven harmonic oscillator). In any of these cases, the stationary-state properties are easily found: due to the bilinear nature of the Hamiltonian and the dissipator, the Wigner density of the system remains Gaussian and its moments are easily found from a closed set of equations of motion. For the nondegenerate PA considered here, this results, for instance, in an occupation
\begin{equation}
\langle n_{a}\rangle_{\mathrm{st}}=\langle n_{b}\rangle_{\mathrm{st}}=\frac{\left(E_{J}/E^{c}_{J}\right)^{2}}{2\left[1-\left(E_{J}/E^{c}_{J}\right)^{2}\right]}\,,
\end{equation}
which diverges once the driving strength reaches the amplification threshold $E^{c}_{J}=(\hbar\sqrt{\gamma_{a}\gamma_{b}}/\sqrt{\kappa_{a}\kappa_{b}})e^{(\kappa_{a}+\kappa_{b})/2}$.
Note that this divergence occurs only here in the simplified linearized model but is cured in the full problem due to higher order terms in the RWA Hamiltonian regularizing the energy gain (cf. Ref.~\cite{Armour2015} for a detailed discussion of the semiclassical dynamics).

For the time-resolved statistics, we follow a similar scheme employing a quantum regression approach \cite{Breuer2002} to set up a closed set of equations for expressions such as $\langle a^{\dagger}(0)a^{\dagger}(\tau)a(\tau)a(0)\rangle_{\mathrm{st}}$ and $\langle b^{\dagger}(0)a^{\dagger}(\tau)a(\tau)b(0)\rangle_{\mathrm{st}}$. We find explicit analytical expressions for the single-resonator second-order correlation function $g^{(2)}_{aa}(\tau)$ and the cross-resonator second-order correlation function $g^{(2)}_{ab}(\tau)$, whose time-dependence are shown in  Fig.~\ref{fig_Fig_2} for various driving strengths (and symmetric decay rates, $r=1$).

The common two-photon excitation process of the PA Hamiltonian leads to a bunching, both for photons leaking from the same cavity with $g^{(2)}_{aa}(0)=2$ (for any driving strength), and more obvious and more pronounced for photons from different cavities with $g^{(2)}_{ab}(0)$ in leading order diverging as $(E_{J}/E^{c}_{J})^{-2}$ for weak driving (cf. the two-photon creation processes within a single cavity discussed in Refs.~\cite{Leppaekangas2013,Gramich2013,Kubala2015,Dambach2015,Padurariu2012}.).

This divergence reflects the simple fact that with a certain fixed probability the first detected photon [within the numerator in the definition of $g^{(2)}(\tau)$] is actually the first photon of a simultaneously created pair of photons to leave the cavity. Hence, it will shortly be followed by the remaining counterpart photon. In consequence for weak driving, the two-photon detection probability for short delay times is proportional to the excitation probability, its long-time limit [at the same time the denominator in  $g^{(2)}(\tau)$] is proportional to $\langle n\rangle_{\mathrm{st}}^{2}$, as the observed photons stem from distinct pairs.

The bunched nature of photon creation thus results in the observed divergence and more generally in the appearance of different timescales. In the waiting-time distribution (not shown), crossover of an exponential decay over a scale associated with the typical duration of a bunch, $\sim 1/\gamma$, and over the typical time between bunches, $\sim 1/\gamma\langle n\rangle_{\mathrm{st}}$, occur. Notably, due to the exclusion of decays during the ``waiting time'', the dynamics becomes non-Gaussian and equations of motion in the calculation of the WTD do not close. 
This non-Gaussian feature inherent in the WTD for a parametric amplifier stands in contrast to the simplest linearized version of Josephson photonics: For the driven harmonic oscillator, which results for the fundamental single-mode resonance, the stationary state is an eigenstate of the jump operator describing decay from the cavity, and the resulting WTD is a trivial exponential decay \cite{Dambach2015}.

\subsection{Two linked two-level systems}
\label{subsec_Two_linked_two-level_systems}

Directly solvable models other than the PA case are realized, when the state space accessible is limited. In Josephson photonics this occurs (within rotating wave approximation and at zero temperature) due to vanishing transition matrix elements of the drive Hamiltonian as discussed above. The simplest of these cases is here achieved for $\kappa_a=\kappa_b=2$, where the cavities are effectively reduced to two linked two-level systems.

Again, it is instructive to highlight similarities and differences to the corresponding one-mode case of a single TLS. Considering stationary properties first, we find for the mean occupation of the TLS a,
\begin{equation}
\langle n_{a}\rangle_{\mathrm{st}}=\frac{(E_{J}/E^{c}_{J})^{2}}{\left[1+(E_{J}/E^{c}_{J})^{2}\right]\left(1+r^{2}\right)}\,,
\label{Eq_two_two-level_systems_occupation}
\end{equation}
where $\langle n_{b}\rangle_{\mathrm{st}}$ follows by replacing $r\rightarrow 1/r$. 

\begin{figure}
\centering\hspace*{-0.45cm}
\includegraphics[width=\columnwidth]{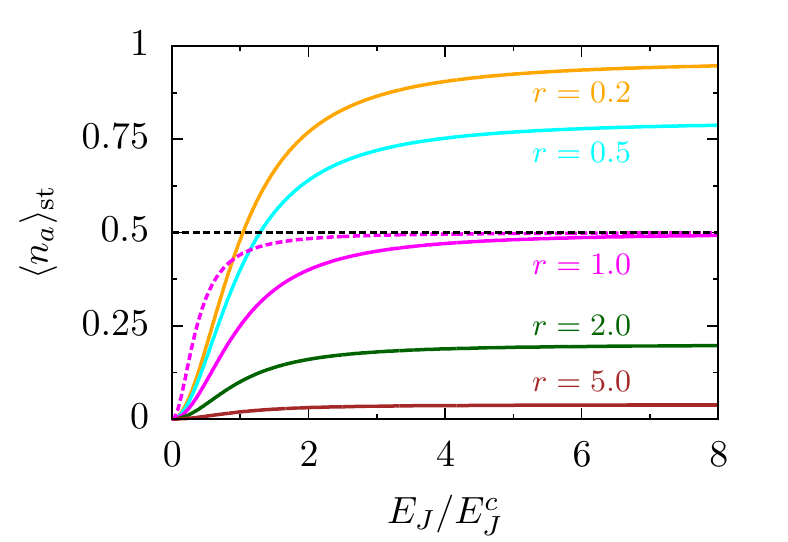}
\caption{Stationary mean photon number $\langle n_{a}\rangle_{\mathrm{st}}$ for two linked two-level systems ($\kappa_{a}=\kappa_{b}=2$) with differing damping, $r\neq1$.
While a single two-level system (dashed) can not be driven to population inversion, the less strongly damped of the two TLSs will reach inversion, $\langle n\rangle_{\mathrm{st}}>0.5$, for sufficiently strong driving. $E_{J}/E^{c}_{J}>1$ is required in the limit $r\rightarrow 0$.}
\label{fig_Fig_3}
\end{figure}

\begin{figure*}
\centering\hspace*{-0.5cm}
\includegraphics[width=1.0\textwidth]{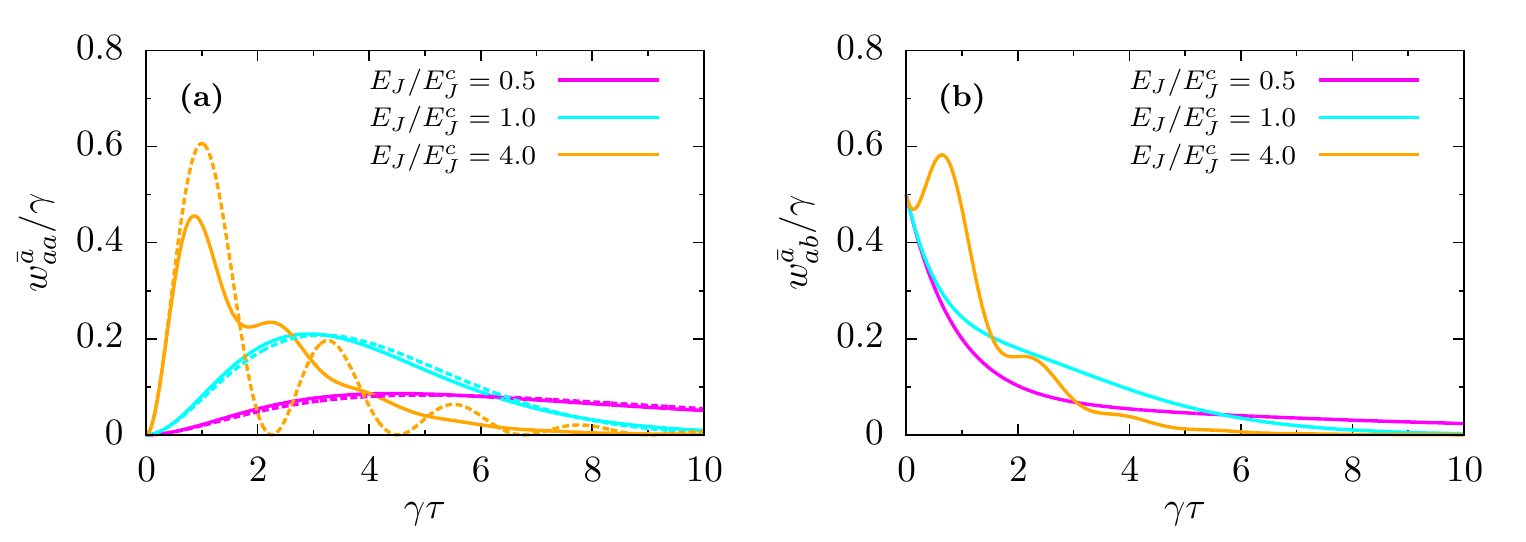}
\caption{Waiting-time distributions (a) $\smash{w^{\bar{a}}_{aa}(\tau)}$ and (b) $\smash{w^{\bar{a}}_{ab}(\tau)}$ for the two linked two-level systems ($\kappa_{a}=\kappa_{b}=2$) and different values of the driving strength $E_{J}/E^{c}_{J}$ for symmetric damping, $r=1$.
The additional dashed lines in (a) refer to the waiting-time distribution $w(\tau)$ of a single two-level system with the identical stationary mean photon number, which shows pronounced Rabi-type oscillations for strong driving.
The presence of a second TLS b has a strong impact on $w^{\bar{a}}_{aa}(\tau)$ in that regime, as (de-)excitation by a tunneling Cooper pair can only happen simultaneously in both TLSs resulting in a more complex dynamics. The time dependence of the cross-resonator WTD, $w^{\bar{a}}_{ab}(\tau)$, in (b) sensitively depends on the driving strength $E_{J}/E^{c}_{J}$.
}
\label{fig_Fig_4} 
\end{figure*}

For any deviation from the strictly symmetric case, $r\neq1$, that cavity which is less strongly damped can achieve occupation inversion, $ \langle n_{a(b)}\rangle_{\mathrm{st}} >1/2$, for sufficiently strong driving, see Fig.~\ref{fig_Fig_3}. While this can not happen for a single TLS, where stimulated absorption and emission from/to the drive field balance in the strong-driving limit, it becomes possible within the larger state space of two TLSs. Then, the doubly-occupied state $\ket{1,1}$ acts like a pump state in a standard lasing scenario. The decay of a photon from double occupation leads to a trapping state blocking both Rabi-type processes in the two-photon drive Hamiltonian, renewed excitation or de-excitation. This trapping is particularly effective for large asymmetry, where it can be exploited for the generation of Fock states \cite{Souquet2016}.

In the time-dependent properties, the partial blockade of Rabi-type oscillations is also apparent. In Fig.~\hyperref[fig_Fig_4]{4(a)} we contrast the distribution of the waiting time between two emission events from cavity a in the two-cavity case, $\smash{w^{\bar{a}}_{aa}(\tau)}$, to the equivalent single-cavity WTD with identical mean photon occupation. The pronounced Rabi-type oscillations observed for strong driving in the single-cavity case are replaced by a more complex pattern. In particular, the partial blocking of oscillations due to the trapping state leads to a vanishing of the recurring ``dark-times'' observed in the single-mode case. The cross-resonator WTD in Fig.~\hyperref[fig_Fig_4]{4(b)} takes a value $\smash{w^{\bar{a}}_{ab}(\tau=0)}=r/(1+r^2)$ independent of driving strength, while its time-dependence is (highly) unusual for certain values of $E_{J}/E^{c}_{J}$. 

Comparing solid and dashed lines in Fig.~\hyperref[fig_Fig_4]{4(a)} demonstrated an impact of the presence of a second TLS b on the individual waiting time between subsequent emissions from TLS a. What is not clear from these results alone, is whether an important property of emission from a single TLS also holds here. That so-called \emph{renewal} property \cite{Cox1995} implies that each photon emission leaves the system in a unique reset state leading to \emph{identical and independent} probability distributions for consecutive events and thus to uncorrelated waiting times.

If renewal theory holds, the two functions $g_{aa}^{(2)}$ and $w_{aa}^{\bar{a}}$ are directly related in Laplace space and provide identical information \cite{Emary2012}. From that one-to-one correspondence, a necessary condition for two Fano-type factors can be deduced \cite{Albert2012}. One Fano factor is defined in the spirit of \emph{full counting statistics}, from the variance of the number of photons $N_{a,T}$ leaked during a long accumulation time T from oscillator a, 
\begin{equation}
\begin{aligned}
F_{a}^{\mathrm{FCS}}&=\frac{\langle N^{2}_{a,\infty}\rangle-\langle N_{a,\infty}\rangle^{2}}{\langle N_{a,\infty}\rangle}\\&=1+2\gamma_{a}\langle n_{a}\rangle_{\mathrm{st}}\int^{\infty}_{0}\mathrm{d}\tau\left[g^{(2)}_{aa}(\tau)-1\right],
\end{aligned}
\end{equation}      
and is directly linked to $g_{aa}^{(2)}(\tau)$ \cite{Mandel1962}.
Another Fano-type factor is defined on the basis of the \emph{first two cumulants of the waiting-time distribution} $w^{\bar{a}}_{aa}(\tau)$:
\begin{equation}
F_{a}^{\mathrm{WTD}}=\frac{\langle\tau_{a}^{2}\rangle-\langle\tau_{a}\rangle^{2}}{\langle\tau_{a}\rangle^{2}}\,.
\end{equation}
If emission events are described by renewal processes, then $F_{a}^{\mathrm{FCS}}=F_{a}^{\mathrm{WTD}}$.

In fact, using analytical results for $\langle n_{a}\rangle_{\mathrm{st}}$, $g^{(2)}_{aa}(\tau)$, and $w^{\bar{a}}_{aa}(\tau)$,
\begin{equation}
\!\:\!\:\!\!\:\:\!\frac{F_{a}^{\mathrm{WTD}}}{F^{\mathrm{FCS}}_{a}}\!\:\!\!-\:\!\!\!1\!\:\!\!=\:\!\!\!\frac{(E_{J}\!/\!E^{c}_{J})^{4}r^{4}/(1\!\:\!\!+\:\!\!\!r^{2})^{2}}{(E_{J}\!/\!E^{c}_{J})^{4}(1\!\:\!\!+\!\!\:\!r^{4})\!\:\!\!-\:\!\!\!2(E_{J}\!/\!E^{c}_{J})r^{2}\!\!\:\!+\!\!\:\!(1\!\!\!\:+\!\!\:\!r^{2})^{2}}\,,\!\!\!\!
\end{equation}
proving that a subsystem of the two linked TLSs has no renewal character, except for the limiting cases of weak driving $E_{J}/E^{c}_{J}\rightarrow 0$ (where individual tunneling processes are uncorrelated) or $r\rightarrow 0$ and $r\rightarrow\infty$. Only in the latter cases, (nearly) every emission from a resets the system to the very same state, while in general renewal theory does not hold and consecutive waiting times are indeed correlated.  

\begin{figure}
\centering
\includegraphics[width=1.0\columnwidth]{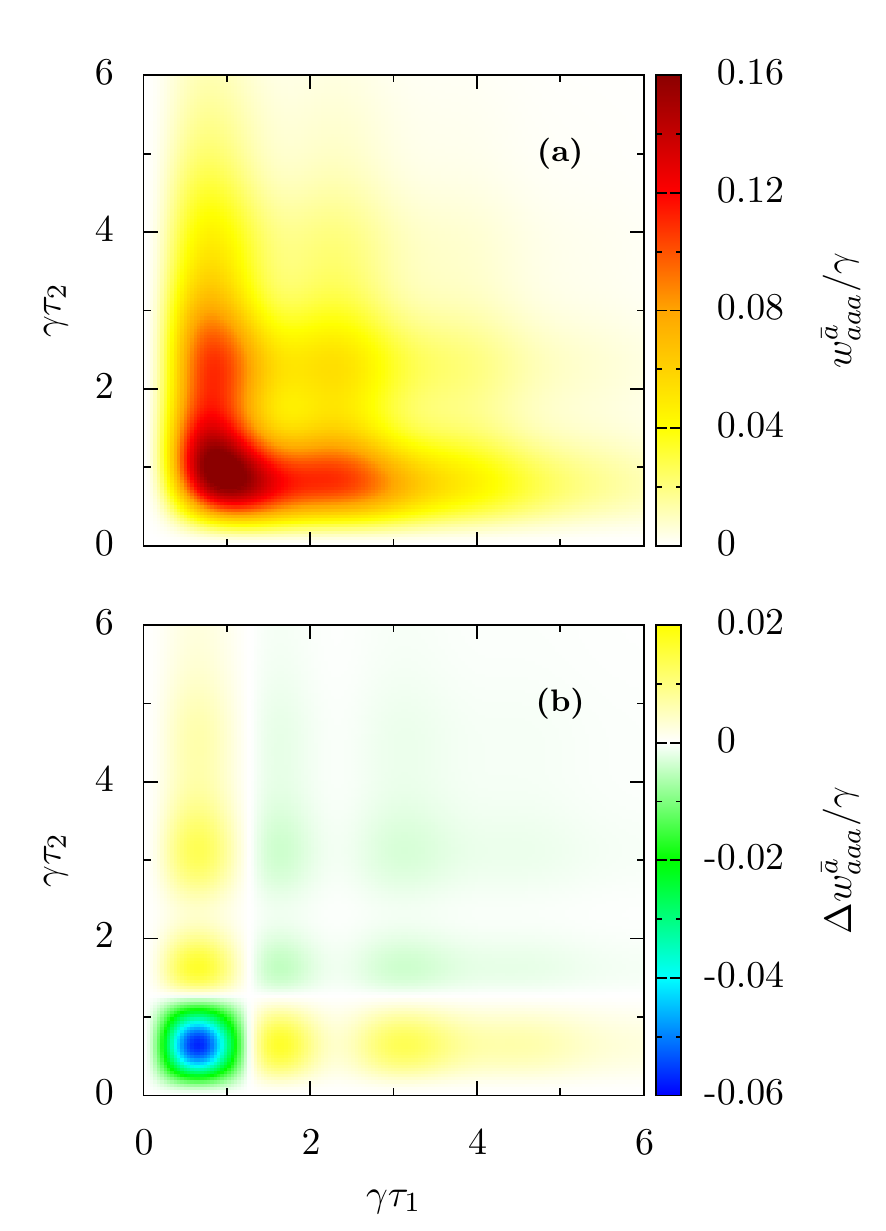}
\caption{(a) Joint waiting-time distribution $w^{\bar{a}}_{aaa}(\tau_{2},\tau_{1})$ for two symmetric TLSs ($\kappa_{a}=\kappa_{b}=2$) and strong driving, $E_{J}/E^{c}_{J}=4.0$. It gives the probability to find the two subsequent photonic waiting times $\tau_{1}$ and $\tau_{2}$. (b) The difference $\Delta w^{\bar{a}}_{aaa}(\tau_{2},\tau_{1})=w^{\bar{a}}_{aaa}(\tau_{2},\tau_{1})-w^{\bar{a}}_{aa}(\tau_{2})w^{\bar{a}}_{aa}(\tau_{1})$ between this joint distribution and its factorized form does not vanish. This implies correlations between subsequent waiting times and, in contrast to a single two-level system, renewal theory thus does not hold. 
}
\label{fig_Fig_5} 
\end{figure}

Such correlations can alternatively be directly revealed employing the concept of \emph{joint waiting-time distributions} \cite{Dasenbrook2015}. These describe the distribution of a given sequence of several consecutive waiting times. For instance, the two-time joint waiting-time distribution 
\begin{equation}
w^{\bar{a}}_{aaa}(\tau_{2},\tau_{1})=\frac{\langle\mathfrak{J}_{a}e^{\mathfrak{L}_{\bar{a}}\tau_{2}}\mathfrak{J}_{a}e^{\mathfrak{L}_{\bar{a}}\tau_{1}}\mathfrak{J}_{a}\rangle_{\mathrm{st}}}{\langle\mathfrak{J}_{a}\rangle_{\mathrm{st}}}
\end{equation}
is the probability distribution to detect two subsequent time delays $\tau_{1}$ and $\tau_{2}$ between emitted photons from resonator a. Displayed in Fig.~\hyperref[fig_Fig_5]{5(a)} for strong driving $E_{J}/E^{c}_{J}=4.0$ and symmetric decay rates, $r=1$, it shows a pronounced maximum close to $\tau_{2}=\tau_{1}=1/\gamma$.

For uncorrelated, statistically independent and equally distributed successive waiting times $\tau_{1}$ and $\tau_{2}$ (as expected in the renewal case), the joint distribution would factorize to the individual distributions and the difference $\Delta w^{\bar{a}}_{aaa}(\tau_{2},\tau_{1})=w^{\bar{a}}_{aaa}(\tau_{2},\tau_{1})-w^{\bar{a}}_{aa}(\tau_{2})w^{\bar{a}}_{aa}(\tau_{1})$, shown in Fig.~\hyperref[fig_Fig_5]{5(b)}, would vanish. 
In fact, we observe waiting times correlated according to a type of ``gambler's fallacy'': a short waiting time is likely followed by a long waiting time and vice versa.

\subsection{Anti-Jaynes-Cummings system}
\label{subsec_Anti-Jaynes-Cummings_system}

\begin{figure*}
\centering
\includegraphics[width=1.0\textwidth]{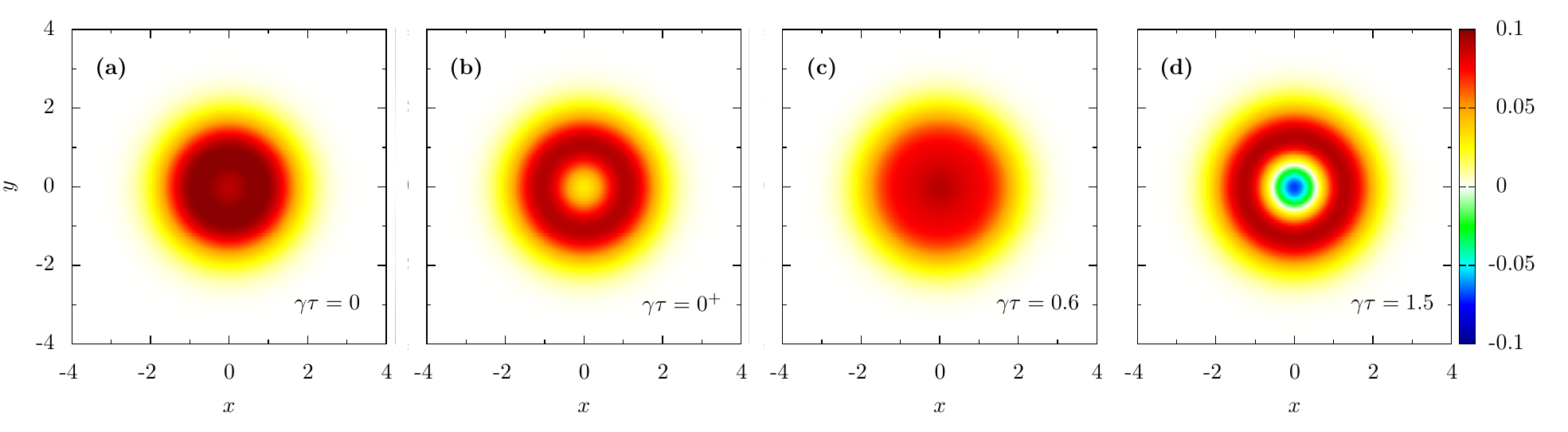}
\caption{Phase-space representation of the harmonic subsystem a in an anti-Jaynes-Cummings configuration ($\kappa_{a}=0.001$, $\kappa_{b}=2.0$) for $E_{J}/E^{c}_{J}=2.0$ and $r=1.0$. 
The TLS b imprints weak non-Gaussian features on the stationary Wigner density $W(x,y)$ (a), which become more pronounced after a photon emission event from the TLS is detected (b). Observing the subsequent time evolution without emissions from a occurring (i.e., under $\mathfrak{L}_{\bar{a}}$), first non-Gaussian features are smeared (c) but later non-classical signatures [i.e., negative values in (d)] develop.}
\label{fig_Fig_6}
\end{figure*}

\begin{figure*}
\centering\hspace*{-0.65cm}
\includegraphics[width=1.0\textwidth]{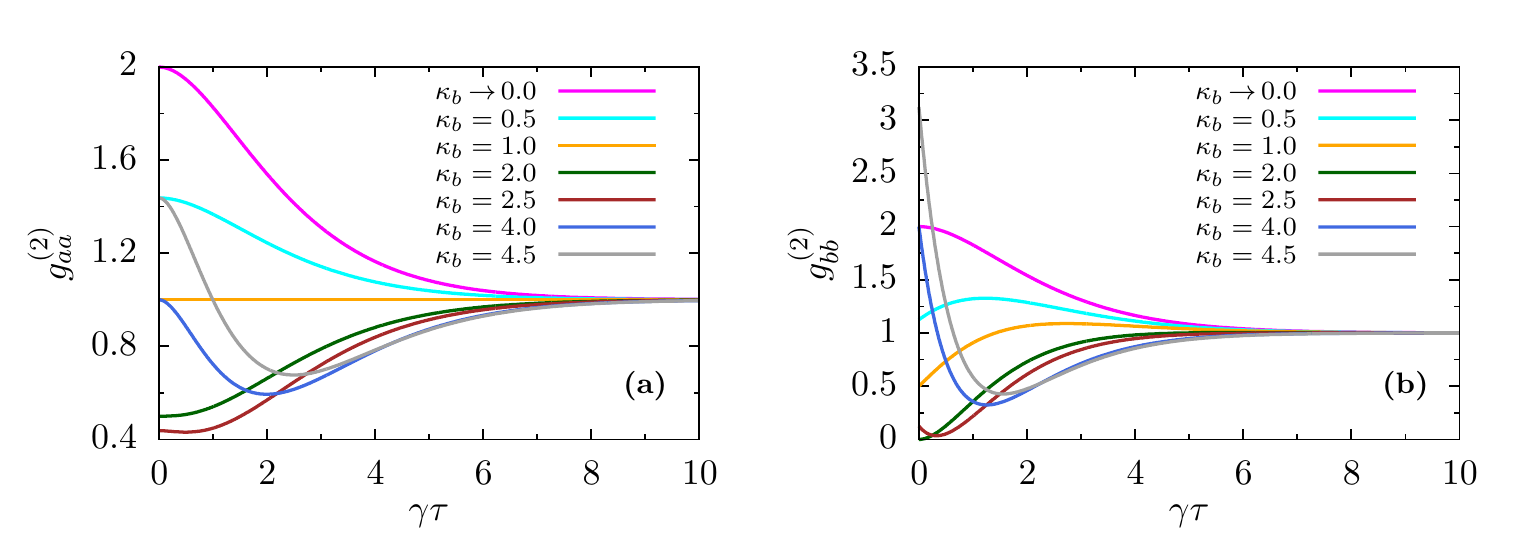}
\caption{Second-order correlation functions (a) $g^{(2)}_{aa}(\tau)$ and (b) $g^{(2)}_{bb}(\tau)$ for $\kappa_{a}\rightarrow0$ and different values of $\kappa_{b}$ in the weak-driving limit $E_{J}/E^{c}_{J}\rightarrow 0$ for symmetric damping rates, $r=1$. Changing  $\kappa_{b}$ tunes the system from a nondegenerate parametric amplifier ($\kappa_{b}\rightarrow0$) to an anti-Jaynes-Cummings system ($\kappa_{b}=2.0$). Depending on $\kappa_{b}$, the photon statistics of the two resonators range from strong bunching to complete antibunching. 
}
\label{fig_Fig_7} 
\end{figure*}

As the last instructive special case, we consider the combination of a harmonic oscillator and a TLS with an anti-JC driving, which (de-)excites both systems simultaneously. This is realized by the two-cavity Josephson-photonics Hamiltonian for the specific values $\kappa_{a}\rightarrow 0$ and $\kappa_{b}=2$.  

The physics of the anti-JC system can also been seen as approximate description of a more generic scenarios of two cavities with highly different $\kappa$: one cavity with small quantum fluctuations behaving only weakly nonlinear, and one cavity with sufficiently large $\kappa$ to restrict excitations within a few-level system.
Due to the presence of (symmetry-breaking) dissipative processes, the damped anti-Jaynes-Cummings system is not trivially linked to the standard Jaynes-Cumming case \cite{Budini2003}. Exact analytical treatments of the damped Jaynes-Cummings system have been performed \cite{Briegel1993,Daeubler1993}, but already the calculation of stationary properties is quite involved. For our studies of the damped anti-JC, we therefore restrict ourselves to numerical results. 

Some interesting insights, however, can already be traced back to the simple equation, $\gamma_{a}\langle n_{a}\rangle_{\mathrm{st}}=\gamma_{b}\langle n_{b}\rangle_{\mathrm{st}}$, balancing the total loss rates from both resonators. 
It basically stems from energy conservation and the simultaneous excitation process of photons and actually holds for any $\kappa_{a(b)}$.
The coupling of the mean occupation of the two cavities leads on the one hand to population inversion in the TLS b, which for the anti-JC system can be reached even for $r<1$ for sufficiently strong driving. On the other hand, it leads to a clear signature of the state-space restrictions of the TLS (or generically a cavity with large $\kappa$) in the occupation as well as in the dynamics of the harmonic resonator a.

This imprinting of the nonlinear quantum character (i.e., of the restricted level spectrum) of subsystem b on the harmonic subsystem a allows for the creation and observation of non-Gaussian or even quantum states in a continuous variable system.

While highly unequal damping and very strong driving enabled Fock-state generation in two TLSs (see above and Ref.~\cite{Souquet2016}), we will here consider moderate driving and equal decay rates. In consequence, the stationary (reduced)  Wigner density of the harmonic subsystem a in Fig.~\hyperref[fig_Fig_6]{6(a)} shows only some weak non-Gaussian features and the slight central dip is far from reaching negative values (and whereby indicating nonclassicality).

If a photon jump out of the TLS b is detected, the central dip in the resulting Wigner function of the harmonic system a immediately after the jump [Fig.~\hyperref[fig_Fig_6]{6(b)}] becomes pronounced. The necessity of an occupied TLS before the jump, also implies less contribution of the ground state of a and clearer signatures of the excited state(s).
In the following time evolution without observing photons from a (i.e., following the spirit of the WTD definition),
we initially observe an approach to a more Gaussian shape, Fig.~\hyperref[fig_Fig_6]{6(c)}, (as it becomes more and more likely that the harmonic oscillator a had already relaxed to its ground state before the initial emission event from b took place). 
For larger times, however, the Wigner function [Fig.~\hyperref[fig_Fig_6]{6(d)}] develops a deep dip reaching negative values, indicating increasing contributions from re-excitation of a and b.

Wigner-density dynamics \-- conditioned on the observation or absence of various emission events \-- can thus be used to distill the nonclassical imprint of the TLS on the harmonic system.

\subsection{Crossover in the weak-driving limit}
\label{subsec_Crossover_in_the_weak-driving_limit}

Numerical results for generic $\kappa_{a(b)}$ parameters are readily available but offer few new insights. Here, we now turn instead to the weak-driving regime, where we can actually gain analytical results for the time-dependent statistics by employing a perturbative approach. In Fig.~\ref{fig_Fig_7}, we show some results of this calculation, namely the time-dependent single-resonator correlations $g^{(2)}_{aa}(\tau)$ and $g^{(2)}_{bb}(\tau)$, illustrating the crossover between the PA and the anti-JC special cases, i.e., we keep cavity a harmonic, $\kappa_a \rightarrow 0$, and vary $\kappa_b$.
The reader will readily recognize some of the results discussed above incorporated in the figure: the $g_{aa(bb)}^{(2)}(\tau=0)$ bunching value of $2$ for the PA case crosses over into an antibunching regime with the characteristic single-photon source value $g^{(2)}_{bb}(\tau=0)=0$ of perfect antibunching of the TLS for $\kappa_b=2$. Note that photons from resonator a, are maximally but not completely antibunched at the slightly higher value $\kappa_{b}=2.5$. Further increasing $\kappa_{b}>2+\sqrt{2}$, zero-time bunching with nonetheless sub-Poissonian Fano factor $F^{\mathrm{FCS}}$ can be observed and regimes, where one cavity acts as a bunched, the other as an antibunched source, can be identified.

\subsection{Conclusions and outlook}
\label{subsec_Conclusions_and_outlook}

Superconducting circuits based on a voltage-biased Josephson junction and one or several series-connected electromagnetic oscillators constitute excellent candidates for designing \emph{versatile sources of quantum microwaves}.
Various basic excitation processes can be selected by the choice of the bias voltage; here, we have concentrated on the resonance, where each tunneling Cooper pair excites a pair of photons, one in each of the two cavities coupled to the junction. 
The impact of the nonlinearity of the Josephson junction on the system's dynamics and the properties of the emitted radiation is governed by the Josephson energy $E_{J}$ determining the overall driving strength and the dimensionless parameter $\kappa_{a(b)}$ characterizing the importance of charge quantization in the circuit. Entering the transition matrix elements of the drive, $\kappa_{a(b)}$ specifies if nonlinearities matter already on the single- or on the many-photon level only. 
Adjusting $\kappa_{a(b)}$ thus allows to find a nondegenerate parametric amplifier, an anti-Jaynes-Cummings system, or two linked two-level systems realized in the system.

We have analyzed second-order correlation functions $g^{(2)}(\tau)$ and waiting-time distributions $w(\tau)$  and found various scenarios, where bunched, antibunched, and other non-classical light sources are attained.
The great variety of highly nonclassical states can be traced back to the complex interplay of the two oscillator subunits, which cannot be found for simple single-mode creation processes.
 
Future work (in preparation) will extend the concepts studied here and employ the correlated excitation process and the resulting correlated emission in a two- or multi-mode setup to generate highly entangled multi-qubit states.

\begin{acknowledgments}
The authors thank A.~D. Armour and F. Portier for valuable discussions. Financial support was provided by the Deutsche Forschungsgemeinschaft (DFG) through Grant No. AN336/6-1 and SFB/TRR21 as well as by the IQST.
\end{acknowledgments}

\bibliography{references}

\end{document}